\documentclass[aps,prb,twocolumn,superscriptaddress]{revtex4}
\usepackage{graphicx}
\usepackage{dcolumn}
\usepackage{amsmath}
\usepackage{amssymb}
\usepackage{subfigure,amsmath,verbatim,moreverb,bm}
\def\be{\begin{equation}}
\def\ee{\end{equation}}
\def\ber{\begin{eqnarray}}
\def\eer{\end{eqnarray}}

\def\av{{\bf a}}

\def\rv{{\bf r}}

\def\Im{\rm Im}
\def\Re{\rm Re}
\begin{document}
\title{Time-dependent current density functional theory on a lattice}
\author{I. V. Tokatly}
\email{Ilya_Tokatly@ehu.es}
\affiliation{Nano-Bio Spectroscopy group and ETSF Scientific Development Centre, 
  Departamento de F\'isica de Materiales, Universidad del Pa\'is Vasco UPV/EHU, E-20018 San Sebasti\'an, Spain}
\affiliation{IKERBASQUE, Basque Foundation for Science, E-48011 Bilbao, Spain}

\date{\today}
\begin{abstract}
A rigorous formulation of time-dependent current density functional theory (TDCDFT) on a lattice is presented. The density-to-potential mapping and the ${\cal V}$-representability problems are reduced to a solution of a certain nonlinear lattice Schr\"odinger equation, to which the standard existence and uniqueness results for nonliner differential equations are applicable. For two versions of the lattice TDCDFT we prove that any continuous in time current density is locally ${\cal V}$-representable (both interacting and noninteracting), provided in the initial state the local kinetic energy is nonzero everywhere. In most cases of physical interest the ${\cal V}$-representability should also hold globally in time. These results put the application of TDCDFT to any lattice model on a firm ground, and open a way for studying exact properties of exchange correlation potentials. 
\end{abstract}
\maketitle

\section{Introduction}

To interpret most experimental results in condensed matter physics, it is sufficient to know the dynamics of a few reduced collective variables, such as the density of particles $n(\rv,t)$ and/or the density of the current ${\bf j}(\rv,t)$. Indeed, in a typical experiment the behavior of a system is controlled/probed by applying external classical electro-magnetic fields, while the density and the current are the physical observables conjugated the scalar $v(\rv,t)$ and the vector ${\bf A}(\rv,t)$ potentials, respectively. On the theoretical side this observation does not seem to be really helpful if the problem is approached with the standard tools of the many-body theory. In any case, one first has to solve the full many-body problem, and only then the required reduced variables are extracted by tracing out all experimentally irrelevant degrees of freedom. Obviously the accurate solution of a nonequilibrium quantum many-body problem for more or less realistic interacting systems is practically impossible. Therefore we have either to rely on some, mostly uncontrollable, approximations within the standard many-body approach, or to look for alternative theoretical schemes. 

One of the most popular alternatives is offered by the time-dependent density functional theory (TDDFT) \cite{TDDFT2006}, which actually states that it is possible, at least in principle, to trace out the irrelevant microscopic degrees of freedom at the very beginning, and to formulate a closed theory that operates only with the observable of interest -- the density $n(\rv,t)$. A theory which considers the current ${\bf j}(\rv,t)$ as a basic collective variable is known as the time-dependent current density functional theory (TDCDFT)\cite{GhoDha1988,Vignale2004}. Apparently a practical application of TDDFT/TDCDFT inevitably involves approximations, but since we are aiming at the reduced description of a system there is a hope that a sensitivity to the accuracy of approximations is somewhat lessened. Unfortunately, in spite of an impressive number of applications, we still do not have a clear methodology for constructing and improving approximations in TDDFT. In fact, we even do not understand why common {\it ad hoc} constructions, like the adiabatic local density approximation, frequently give good results (it is much easier to understand why they fail). Probably one of the reasons for this situation is that the formal mathematical justification of TDDFT/TDCDFT, as well as the understanding of fundamental limitations of this approach, if there are any, is still far from being complete. 

Presently the justification of TDDFT relies on the Runge-Gross mapping theorem \cite{RunGro1984}, which states that for a given initial many-body state $\psi_0$, there is a one-to-one correspondence between the density $n(\rv,t)$ and the external potential $v(\rv,t)$ driving the dynamics. This implies that the external potential, the many-body wave function, and thus any observable are unique functionals of the density, which indicates a possibility to formulate a closed theory of a single collective variable, $n(\rv,t)$. However, Runge and Gross have proved the above mapping only for potentials which are analytic in time $t$ ($t$-analytic) around the initial point $t_0$. Therefore many physically eligible potentials are formally excluded from the scope of TDDFT. There are indications that the $t$-analyticity is not a fundamental limitation of the theory as in the linear response regime the mapping can be proved for a much wider class of Laplace transformable potentials \cite{vanLeeuwen2001}. Nonetheless, beyond the linear response, a confirmation of the Runge-Gross theorem for a more general class of potentials is still lacking. The situation with the current-to-vector potential mapping in TDCDFT \cite{GhoDha1988} is exactly the same.

A more serious point is that the mapping theorem, in its standard form, is not sufficient to formally justify most of the practical applications of TDDFT/TDCDFT. The unique density-to-potential mapping is proved only for $v$-representable densities, i.~e. for the densities generated by the Schr\"odinger dynamics in the presence of some external potential. On the other hand, the applications almost always employ the Kohn-Sham (KS) formalism. The KS construction assumes that, by properly adjusting a potential (so called KS potential) in a fictitious system of noninteracting KS particles, one can exactly reproduce the physical density $n(\rv,t)$ of the interacting system. In other words, the applicability of the KS formalism requires that the ``interacting $v$-representable'' density is also a ``noninteracting $v$-representable''. This is not guaranteed by the Runge-Gross theorem even for $t$-analytic potentials. The generalization of the TDDFT mapping theorem by van~Leeuwen \cite{vanLeeuwen1999} (later extended by Vignale to the TDCDFT case \cite{Vignale2004}) is aimed at solving the $v$-representability problem. Van~Leeuwen has shown \cite{vanLeeuwen1999} that, assuming the $t$-analyticity both for the potential $v(\rv,t)$ and for the density $n(\rv,t)$ in the interacting system, one can uniquely construct all coefficients in the $t$-power series expansion of the KS potential \cite{vanLeeuwen-note} around the initial time $t_0$. Note that the $t$-analyticity of the density is a much more severe restriction than the $t$-analyticity of the potential employed by Runge and Gross. One can easily find many explicit examples where the density, evolving under an absolutely physical, $t$-analytic, and infinitely smooth in space potential, is not $t$-analytic, and can, therefore, not be represented by a uniformly convergent Taylor series \cite{Maitra2010}. A closely related drawback is that the existence of all coefficients in the power series expansion \cite{RugPenBau-note} for the KS potential does not imply that this series converges, a situation well known in the theory of differential equations. Probably, under some, most likely very strong, restrictions the power series approach is valid, but the range of validity remains unclear.

In general the state-of-the-art situation in the field of foundations of TDDFT/TDCDFT can be viewed as follows. From a mathematical purist point of view, the general validity of this theory is a plausible and useful conjecture, still waiting for a rigorous justification. From the more soft and practical point of view of a physicist, one of the main problems is that the common way of stating and proving the theorems in TDDFT/TDCDFT is not really helpful in understanding the general properties of the theory and the key functionals that we have to approximate in practice. The Runge-Gross theorem only says that the functional $v[n]$ is unique, but it does not give any idea of where it comes from and how it may look like. The van~Leeuwen procedure is more constructive, but the power series structure, together with a highly implicit and complicated form of the coefficients, makes the analysis of a solution practically impossible. Hence both mathematical and, to some extent, physical problems are eventually connected to the fact that the power series expansion is internally built in the very idea of all existing proofs \cite{RunGro1984,GhoDha1988,vanLeeuwen1999,Vignale2004}. Therefore, to make further progress, it is desirable to find an alternative statement of the problem.

Recently it has been realized that the problem of the existence of TDDFT/TDCDFT can be indeed formulated differently. Namely, the density-to-potential mapping and the $v$-reprsentability problems can be restated in the form of the existence and uniqueness of solutions to a certain universal nonlinear Schr\"odinger equation (NLSE). In brief, the appearance of NLSE in TDDFT/TDCDFT can be described as follows. Given the initial state $\psi_0$ and the external (scalar or vector) potential ${\cal V}(t)$, the standard Schr\"odinger equation determines the wave function $\psi(t)$ and the corresponding density ${\cal N}(t)$ as unique functionals of $\psi_0$ and ${\cal V}(t)$. TDDFT does exist if the inverse problem of reconstructing the wave function and the potential from a given density has a unique solution. It turns out that this inverse problem corresponds to the solution of a special time-dependent NLSE, which, for a given initial state $\psi_0$ and the density ${\cal N}(t)$, returns the functionals $\psi[\psi_0,{\cal N}](t)$ and ${\cal V}[\psi_0,{\cal N}](t)$. This approach has been first formulated in the context of the time-dependent deformation functional theory (TDDefFT)\cite{TokatlyPRB2007,TokatlyPCCP2009}, and, more recently, a similar formulation has been proposed for TDDFT \cite{Maitra2010}. The restatement of the problem in the form of NLSE clearly demonstrates the origin of the universal functionals in TDDFT -- they come about naturally as solutions to a well defined differential equation. It also does not rely on the power series expansion, although the latter can be used as the simplest tool for proving the uniqueness of a solution if the $t$-analyticity is taken for granted \cite{TokatlyPRB2007,TokatlyPCCP2009}. It should be noted that up to now the new setup has been analyzed only within the standard assumptions of $t$-analyticity. However, because of a more general and clear statement of the problem, there is hope that on this way a proper mathematical rigor and better physical understanding can be achieved.

In the present paper I demonstrate how the above ideas may work in practice. The main result of this work is a generalization of the NLSE approach to a lattice many-body theory, and a rigorous formulation of the corresponding lattice TDCDFT. While preserving all essential physics, the lattice version is substantially simpler technically. In this case the universal NLSE reduces to a system of nonlinear ordinary differential equations (ODE), for which many general uniqueness and existence results are known (see, e.~g., Ref.~\onlinecite{CoddingtonODE}). I will show that the discrete NLSE for the lattice TDCDFT can be brought to the form which guaranties that the Picard iterations do converge to a unique solution, provided the current is a continuous function of $t$, and the initial state satisfies certain well defined conditions.  It is known that on a lattice not all time-dependent densities/currents are ${\cal V}$-representable \cite{Baer2008,LiUll2008,Verdozzi2008}. In the present formalism the origin of such potentially dangerous situations becomes especially clear, and we will see that they do not lead to any serious complication in the theory. It is worth noting that standard power series construction, being applied to the TDCDFT on a lattice \cite{StePerCin2010}, suffers all common problems with the assumption of $t$-analyticity and unproved convergence of the series (in fact, we know that in a lattice theory the series should not converge in some cases\cite{Baer2008,LiUll2008,Verdozzi2008}).  In contrast, using the NLSE setup we will rigorously solve the ${\cal V}$-representability problem, and thus proveing the existence of TDCDFT for all lattice many-body systems, e.~g., for various versions of the Hubbard model. Therefore the situation with the time-dependent ${\cal V}$-representability on a lattice turns out to be qualitatively similar to that in the ground state lattice DFT\cite{Kohn1983}.

The structure of the paper is the following. In Sec.~II the general many-body theory on a lattice is reviewed. The aim of this section is to fix the notations, and to introduce basic equations and the physical observables of interest. In Sec.~III I formulate a ``generalized lattice TDCDFT'' which, being very simple conceptually, clearly illustrates main ideas of the NLSE approach. Here the complex hopping parameter plays a role of the external driving one-body potential, while the conjugated ``complex current'' is considered as a basic collective variable. Despite this theory may look a bit unusual physically, it is actually more close mathematically to the TDCDFT in a continuum as it does not have specific lattice non-${\cal V}$-representabilities discussed in Refs.~\onlinecite{Baer2008,LiUll2008,Verdozzi2008}. In Sec.~IV the direct, physical lattice analog of the standard continuum TDCDFT is considered. In this section I formulate and prove the existence and uniqueness theorem for the lattice TDCDFT (Theorem~2), which is probably of main interest for practical applications. The general discussion of the results is given in the Conclusion.

\section{Many-body theory on a lattice}

Let us consider a system of $N$ quantum particles living on a lattice that consists of $M$ discrete sites with the nearest-neighbor hopping. The state of the system at time $t$ is described by a many-body wave function $\psi(\rv_1,\rv_2\dots\rv_N;t)$, where $\rv_j$ ($j=1\dots N$) are the discrete coordinates of particles, which take values on the lattice sites. The dynamics of the system is governed by the following discrete (tight binding) version of the Schr\"odinger equation
\begin{widetext}
\begin{equation}
\label{SE1}
i\partial_t\psi(\rv_1\dots\rv_N;t) = -\sum_{j=1}^{N}\sum_{\av}T_0e^{iA(\rv_j,\rv_j+\av;t)}\psi(\dots \rv_j+\av \dots;t)
+ \sum_{j=1}^{N}\varepsilon(\rv_j;t)\psi(\rv_1\dots\rv_N;t) + \sum_{i>j}V_{\rv_i-\rv_j}\psi(\rv_1\dots\rv_N;t)
\end{equation}
\end{widetext}
where $\av$ are vectors connecting a given site with its nearest neighbors, $T_0$ is the modulus of the hopping parameter, $\varepsilon(\rv;t)$ is the time-dependent on-site energy (the scalar potential on a lattice), $A(\rv,\rv+\av;t)$ is the vector potential (the so called Peierls phase) on a link $\{\rv,\rv+\av\}$, and $V_{\rv-\rv'}$ is the potential of a pairwise interaction between particles. The first term in the right hand side in Eq.~(\ref{SE1}) can be viewed as a finite difference representation of the Laplace operator. It contains summation over all configurations obtained from $N$ particles occupying sites $\rv_1\dots\rv_N$ after a hopping of one of the particles to all possible nearest sites. The modulus of the hopping parameter controls the rate of these hops, while its phase (the vector potential) describes a phase accumulated when a particle is transported along the link. Note that the vector potential should be a skew-symmetric matrix $A(\rv,\rv')=-A(\rv',\rv)$ to ensure the hermiticity of the Hamiltonian in Eq.~(\ref{SE1}). Everywhere below we assume that this condition is fulfilled. Equation (\ref{SE1}) has to be supplemented with the initial condition
\begin{equation}
 \label{init-cond}
\psi(\rv_1\dots\rv_N;t_0) =\psi_0(\rv_1\dots\rv_N)
\end{equation}
The initial state is assumed to be normalized to unity. Since the Hamiltonian in Eq.~(\ref{SE1}) is Hermitian the normalization is preserved during the evolution, i.~e. at any instant $t$ the following normalization condition is fulfilled
\begin{equation}
 \label{norm}
\sum_{\rv_1\dots\rv_N}|\psi(\rv_1\dots\rv_N;t)|^2=1
\end{equation}

It is convenient to eliminate the on-site energies by the gauge transformation
\begin{equation}
 \label{gauge}
\psi(\rv_1\dots\rv_N;t)\to \psi(\rv_1\dots\rv_N;t)e^{-i\sum_{j=1}^N\int_{t_0}^{t}\varepsilon(\rv_j;t')dt'},
\end{equation}
and to redefine the vector potential on a link as follows
\begin{equation}
 \label{A-def}
A(\rv,\rv+\av)\to A(\rv,\rv+\av)-\int_{t_0}^{t}[\varepsilon(\rv;t')-\varepsilon(\rv+\av;t')]dt',
\end{equation}
which corresponds to the description of the external driving field in a temporal gauge. As a result Eq.~(\ref{SE1}) reduces to the form
\begin{widetext}
\begin{equation}
 \label{SE2}
i\partial_t\psi(\rv_1\dots\rv_N;t) = -\sum_{j=1}^{N}\sum_{\av}T(\rv_j,\rv_j+\av;t)\psi(\dots \rv_j+\av\dots;t)
+ \sum_{i>j}V_{\rv_i-\rv_j}\psi(\rv_1\dots\rv_N;t),
\end{equation}
\end{widetext}
where the complex hopping parameter $T(\rv,\rv+\av;t)$ is defined as follows
\begin{equation}
 \label{T-def}
T(\rv,\rv+\av;t) = T_0e^{iA(\rv,\rv+\av;t)}.
\end{equation}
In the latice theory the vector potential $A(\rv,\rv+\av)$ on a link is defined modulo $2\pi$. All vector potential of the form $A(\rv,\rv+\av)+2\pi l$ with integer $l$, being physically identical, belong to the same equivalence class. Below all statements regarding the uniqueness and/or existence of vector potentials always refer to the corresponding equivalence classes.

In general Eq.~(\ref{SE2}) corresponds to a system of $M^N$ ordinary differential equations (ODE) for $M^N$ functions of $t$. In fact, $N_{\cal H}=M^N$ is a dimension of a Hilbert space for $N$ particles on $M$ sites. If the particles are identical the number of independent equations/functions is reduced. For example, for spinless fermions the number of independent equations equals to $\frac{M!}{(M-N)!N!}$, while for $N$ bosons it is $\frac{(M+N-1)!}{(M-1)!N!}$. To simplify some notations, in the following I assume that the particles are identical, but all results are valid without any restriction on the permutation symmetry of the wave functions.

To identify basic physical variables of the lattice TDCDFT we consider the equation of motion for the density $n(\rv;t)$ that is the number of particles on site $\rv$
\begin{equation}
 \label{n-def}
n(\rv;t) = N\sum_{\rv_2\dots\rv_N}|\psi(\rv,\rv_2\dots\rv_N;t)|^2
\end{equation}
By differentiating Eq.~(\ref{n-def}) with respect to time, and using Eq.~(\ref{SE2}) we find
\begin{equation}
 \label{continuity}
\partial_t n(\rv;t) = i\sum_{\av}[Q(\rv,\rv+\av;t) - Q^*(\rv,\rv+\av;t)].
\end{equation}
Here the summation is over all links connected to the site $\rv$, and $Q(\rv,\rv+\av;t)$ is defined as follows
\begin{equation}
 \label{Q-def}
Q(\rv,\rv+\av;t) = T(\rv,\rv+\av;t)\rho(\rv,\rv+\av;t),
\end{equation}
where $\rho(\rv,\rv+\av;t)$ is a ``link density'' (a density matrix on the link $\{\rv,\rv+\av\}$ of the lattice):
\begin{equation}
 \label{rho-def}
\rho(\rv,\rv+\av) = N\sum_{\rv_2\dots\rv_N}\psi^*(\rv,\rv_2\dots\rv_N)\psi(\rv+\av,\rv_2\dots\rv_N).
\end{equation}
Equation (\ref{continuity}) is a lattice version of the usual continuity equation. The quantity $Q(\rv,\rv+\av;t)$ defined on each link of the lattice will play a key role in the formulation of the lattice TDCDFT. It is natural to call $Q(\rv,\rv+\av;t)$ a ``complex link current''. The imaginary part of $Q(\rv,\rv+\av;t)$, entering the continuity equation (\ref{continuity}), is equal to the physical current $J(\rv,\rv+\av;t)$ on a link,
\begin{equation}
 \label{J-def}
J(\rv,\rv+\av;t) = 2\Im Q(\rv,\rv+\av;t),
\end{equation}
which is a flow of particles from the site $\rv$ to the site $\rv+\av$. The real part of the complex link current determines the local kinetic energy $K(\rv,\rv+\av;t)$ on the link
\begin{equation}
 \label{K-def}
K(\rv,\rv+\av;t) = 2\Re Q(\rv,\rv+\av;t)
\end{equation}
Apparently the physical link current is antisymmetric, $J(\rv,\rv')=-J(\rv',\rv)$, the link kinetic energy is symmetric, $K(\rv,\rv')=K(\rv',\rv)$, and the complex current is Hermitian, $Q(\rv,\rv')=Q^*(\rv',\rv)$, under interchanging the link's end points (reversal of the link direction). Finally, using the definition of the physical link current, Eq.~(\ref{J-def}), we can represent the continuity equation (\ref{continuity}) in the form 
\begin{equation}
 \label{continuity2}
\partial_t n(\rv;t) = -\sum_{\av}J(\rv,\rv+\av;t),
\end{equation}
which shows that the sum of link currents flowing away from site $\rv$ equals to the rate of the density decrease on that site. 

\section{Generalized TDCDFT}

In this section I formulate a ``generalized TDCDFT''. That is the theory in which the complex link current $Q(\rv,\rv+\av;t)$ plays a role of the basic collective variable. Our starting point is the Schr\"odinger equation (\ref{SE2}) supplemented with the initial condition of Eq.~(\ref{init-cond}). We consider quantum many-body dynamics driven by a most general one-body external ``potential'' on a latice -- a full complex hopping parameter $T(\rv,\rv+\av;t)$ with time-dependent modulus and phase. Obviously, the complex link current is an observable conjugated to the hoping parameter, i.~e. $Q(\rv,\rv+\av)$ is the variational derivative of the Hamiltonian with respect to $T(\rv,\rv+\av)$.

The standard statement of the problem in quantum mechanics is the following. For a given set $T=\{T(\rv,\rv+\av;t)\}$ of continuous in time ($t$-continuous) complex hopping parameters, the initial value problem of Eqs.~(\ref{SE2}) and (\ref{init-cond}) uniquely determines the many-body wave function $\psi[T,\psi_0](t)$ at all $t>t_0$ as a functional of $T$ and $\psi_0$. Using this wave function we can calculate any observable, in particular, the complex link currents $Q[T,\psi_0](\rv,\rv+\av;t)$. The currents obtained via this procedure we call $T$-representable. 

The generalized TDCDFT exists if the inverse problem possesses a unique solution. Assume that we are given a set of link currents $Q=\{Q(\rv,\rv+\av;t)\}$ and the initial state $\psi_0$. Is it sufficient to uniquely determine the wave function $\psi[Q,\psi_0](t)$, and to reconstruct the hopping parameters $T[Q,\psi_0](\rv,\rv+\av;t)$ which give rise to the prescribed currents? If yes, are there any restrictions on the currents which guarantee such a reconstruction (i.~e. guarantee that the set $Q$ is $T$-representable)? In the standard TDDFT terminology the former question is commonly referred to as the mapping problem, while the latter is the essence of the ${\cal V}$-representability problem. For the generalized lattice TDCDFT the rigorous answers to the above questions can be formulated in a form of the following theorem.

{\it Theorem 1. (Generalized TDCDFT)} --- Let the complex link currents $Q(\rv,\rv+\av;t)$ be continuous functions of $t$, and the initial state $\psi_0$ is such that for all links $|\rho(\rv,\rv+\av;t_0)|\ne 0$. Then,

\noindent
(i) The time-dependent many-body wave function $\psi(t)$ is a unique, retarded functional of $Q(\rv,\rv+\av;t)$ and $\psi_0$;

\noindent
(ii) There is a bijective mapping $Q\leftrightarrow T$ between the complex link currents $Q=\{Q(\rv,\rv+\av;t)\}$ and the complex hoping integrals $T=\{T(\rv,\rv+\av;t)\}$;

\noindent
(iii) The statements (i) and (ii) either hold infinitely long in time, or break down at some finite $t^*>t_0$ if and only if at least one link density vanishes as $t\to t^*$, i.~e., $|\rho(\rv,\rv+\av;t^*)|=0$, which coresponds to $|T(\rv,\rv+\av;t^*)|\to\infty$.

{\it Proof} --- For given complex link currents, we invert Eq.~(\ref{Q-def}) to express the hopping parameters $T(\rv,\rv+\av;t)$ in terms of $Q(\rv,\rv+\av;t)$ and the many-body wave function $\psi(\rv_1\dots\rv_N;t)$ as follows
\begin{equation}
 \label{T(Q)}
T(\rv,\rv+\av;t) = \frac{Q(\rv,\rv+\av;t)}{\rho(\rv,\rv+\av;t)},
\end{equation}
where $\rho(\rv,\rv+\av;t)$ is defined by Eq.~(\ref{rho-def}). Inserting Eq.~(\ref{T(Q)}) into Eq.~(\ref{SE2}) we get the following nonlinear lattice Schr\"odinger equation (NLSE) 
\begin{widetext}
\begin{equation}
 \label{NLSE1}
i\partial_t\psi(\rv_1\dots\rv_N;t) = -\sum_{j=1}^{N}\sum_{\av}\frac{Q(\rv_j,\rv_j+\av;t)}{\rho(\rv_j,\rv_j+\av;t)}
\psi(\dots \rv_j+\av\dots;t) + \sum_{i>j}V_{\rv_i-\rv_j}\psi(\rv_1\dots\rv_N;t).
\end{equation}
To clearly see the structure of the nonlinearity we substitute the definition of the link density, Eq.~(\ref{rho-def}), into the above NLSE, and rewrite it in the following, more explicit form:
\begin{equation}
\label{NLSE1a}
i\partial_t\psi(\rv_1\dots\rv_N;t) = -\sum_{j=1}^{N}\sum_{\av}\frac{Q(\rv_j,\rv_j+\av;t)\psi(\dots \rv_j+\av\dots;t)}{N\sum_{\rv'_2\dots\rv'_N}\psi^*(\rv_j,\rv'_2\dots\rv_N;t)\psi(\rv_j+\av,\rv'_2\dots\rv'_N;t)}
 + \sum_{i>j}V_{\rv_i-\rv_j}\psi(\rv_1\dots\rv_N;t).
\end{equation}
\end{widetext}
Equation (\ref{NLSE1a}) and the initial condition (\ref{init-cond}) constitute a universal problem, which determines the many-body wave function $\psi(t)$ for a given set of complex link currents. Formally we have a Cauchy problem for a system of $N_{\cal H}$ nonliner ODE of the following general form 
\begin{equation}
 \label{abstract-NLSE}
\dot{\bm\psi} = {\bf F}({\bm\psi},t), \qquad {\bm\psi}(t_0)={\bm\psi}_0,
\end{equation}
where ${\bm\psi}\in{\cal H}$ is a $N_{\cal H}$-dimensional vector, and the right hand side is a nonlinear function of $\psi$-variables and time, with the explicit $t$-dependence determined solely by the complex link currents [see Eqs.~(\ref{NLSE1}) or (\ref{NLSE1a})]. 

Let $\Omega$ be a subset of ${\cal H}$, which is defined by the inequalities $|\rho(\rv,\rv+\av)|>0$ for all links. In other words, for any state ${\bm\psi}\in\Omega$ the link densities on all links are nonzero, which ensures nonvanishing denominators in Eq.~(\ref{NLSE1}). Hence if ${\bm\psi}\in\Omega$, then the right hand side in Eq.~(\ref{NLSE1}) is a continuously differentiable function of $\psi$-variables, and therefore, it is locally Lipshitz in $\Omega$. Since by the assumption of the theorem the initial state ${\bm\psi}_0\in\Omega$, the standard uniqueness and existence results for first order ODE are directly applicable. In particular, by the Picard-Lindel\"of theorem (see, e.g., Ref.~\onlinecite{CoddingtonODE}) there exists a finite interval $t_0-\delta<t<t_0+\delta$, with $\delta>0$, where Eq.~(\ref{NLSE1}) has a unique solution, which defines the many-body wave function $\psi[Q,\psi_0](t)$ as a unique functional of the complex link currents and the initial state. Inserting this solution into Eq.~(\ref{T(Q)}) we construct a unique map $Q\mapsto T$. The existence and uniqueness of the inverse map $T\mapsto Q$ is a trivial consequence of the standard linear Schr\"odinger dynamics governed by Eq.~(\ref{SE2}) with given hopping parameters. This proves statements (i) and (ii) locally in time.

The existence and uniqueness theorem for the first order ODE also implies that the local solution can be extended to its maximal existence time which can be infinite (i.~e. the solution is global) or finite. In general, if the maximal existence time $t^*$ is finite, there are only two possible types of behavior at $t\to t^*$. First, the solution becomes unbounded $\|{\bm\psi}\|\to\infty$ as $t\to t^*$ or, second, it reaches the boundary of the subset $\Omega$ where the right hand side of the equation is defined. In our case the first possibility is excluded as the wave function is normalized and thus always bounded. Hence the maximal existence and uniqueness time can be finite only if at $t\to t^*$ at least one link density vanishes, $|\rho(\rv,\rv+\av;t^*)|=0$. Otherwise the solution is necessarily global. This proves the statement (iii) and completes the proof of the theorem.

It is instructive to illustrate the statements of Theorem~1 for a simple explicit example of one particle on a lattice. In the one-particle case the wave function $\psi(\rv;t)$ depends only on one coordinate, while the link density reduses to a simple product $\rho(\rv,\rv+\av)=\psi^*(\rv)\psi(\rv+\av)$. Therefore Eq.~(\ref{NLSE1}) takes the following exactly solvable form
\begin{eqnarray}
 \nonumber
i\partial_t\psi(\rv;t) &=& -\sum_{\av}\frac{Q(\rv,\rv+\av;t)}{\psi^*(\rv;t)\psi(\rv+\av;t)}\psi(\rv+\av;t)\\
\label{NLSE-1part}
&\equiv& -\frac{\sum_{\av}Q(\rv,\rv+\av;t)}{\psi^*(\rv;t)}.
\end{eqnarray}
By integrating this equation with the initial condition $\psi(\rv;t_0)=|\psi_0(\rv)|e^{i\chi_0(\rv)}$ we get the following result for the modulus and the phase of the time-dependent wave function, $\psi(\rv;t)=|\psi(\rv,t)|e^{i\chi(\rv,t)}$,
\begin{eqnarray}
 \label{modPsi-1part}
|\psi(\rv,t)| &=& \sqrt{|\psi_0(\rv)|^2 - \int_{t_0}^{t}\sum_{\av}J(\rv,\rv+\av;t')dt'}, \\
\label{phase-1partQ}
\chi(\rv,t) &=& \chi_0(\rv) + \int_{t_0}^{t}\frac{\sum_{\av}K(\rv,\rv+\av;t')}{2|\psi(\rv,t')|^2}dt',
\end{eqnarray}
where $J(\rv,\rv+\av;t)$ and $K(\rv,\rv+\av;t)$ are defined by Eqs.~(\ref{J-def}) and (\ref{K-def}). 

Equation (\ref{modPsi-1part}) explicitly demonstrates two possibilities indicated in the Theorem~1. The interval of existence is finite if, at least for one site, there is a solution $t^*$ to the equation
\begin{equation}
 \label{t^*}
|\psi_0(\rv)|^2 - \int_{t_0}^{t^*}\sum_{\av}J(\rv,\rv+\av;t')dt'=0.
\end{equation}
This equation means that the number of particles which have left the site $\rv$ by the time $t^*$ (the second term) is equal to the initial occupation of this site (the first term). In other words, at time $t^*$ the site $\rv$ becomes empty and the dynamics stops because it is not anymore possible to support the prescribed link current by adjusting the hopping parameter. If Eq.~(\ref{t^*}) is never fulfilled, the occupation numbers are always finite and the solution of Eqs.~(\ref{modPsi-1part}), (\ref{phase-1partQ}) is global. According to the Theorem~1 this behavior is generic for an arbitrary many-body system. It should be emphasized that the finite interval of existence is guarantied for any set of continuous complex link currents, i.~e. any complex link current is locally $T$-representable.

Obviously, all statements of the above theorem do not depend on a form of the interaction potential. Therefore it is possible to reproduce the same set of complex link currents in different many-body systems with different particle-particle interactions. In particular, this implies the existence of the KS system. Given a physical interacting system, one can always construct a fictitious system of noninteracting KS particles with exactly the same complex current. The only requirement is that the initial states of the interacting and the KS systems should be in the ``$T$-representability subset'' $\Omega$ of the Hilbert space ${\cal H}$. If this requirement is fulfilled, then any set of $t$-continuous complex link currents is both interacting and noninteracting $T$-representable.

\section{Lattice TDCDFT}

A practically unconditional $T$-representability is the main technical advantage of considering the complex link current as the basic variable. In the generalized TDCDFT the prescribed values of $Q(\rv,\rv+\av;t)$ are reproduced by varying both the phases (the vector potential) and the moduli (the hopping rate) of the hopping parameters $T(\rv,\rv+\av;t)$. Because in the generalized setting the hopping rate is allowed to vary, there are no physical restrictions on the time derivative of the density, and therefore the non-${\cal V}$-representable situations discussed in Refs.~\onlinecite{Baer2008,LiUll2008,Verdozzi2008} are absent by construction.

Let us now analyze a more restricted formulation of the lattice TDCDFT, which is a direct lattice analog of the usual TDCDFT in the continuum. Namely, I consider the physical link current $J(\rv,\rv+\av;t)$ as the basic variable and require the hopping parameters $T(\rv,\rv+\av;t)$ to be of the general ``physical'' form, Eq.~(\ref{T-def}). That is, the modulus of the hopping parameter is the same for all links, and fixed as follows
\begin{equation}
 \label{T-modulus}
|T(\rv,\rv+\av;t)|=T_0.
\end{equation}
Only its phase, describing the external vector potential, is allowed to vary in time and from link to link. In this case the link vector potential and the physical link current constitute a pair of conjugated variables -- the current $J(\rv,\rv+\av)$ is a variational derivative of the Hamiltonian with respect to the vector potential $A(\rv,\rv+\av)$. 

Given a set of vector potentials $A=\{A(\rv,\rv+\av;t)\}$, one can solve the initial value problem of Eqs.~(\ref{SE2}), (\ref{T-def}), and (\ref{init-cond}) to get the many-body wave function $\psi[A,\psi_0](t)$ and to construct a unique map $A\mapsto J$ from $A$ to the link currents $J=\{J(\rv,\rv+\av;t)\}$. Such currents we will call $A$-representable. 

The lattice TDCDFT does exist if the inverse problem is well posed. This means that a given current $J$ uniquely determines the wave function $\psi[J,\psi_0](t)$ and the vector potential $A$ which produces that current. To derive NLSE corresponding to this inverse problem, we use Eq.~(\ref{T(Q)}) together with the constraint of Eq.~(\ref{T-modulus}). This allows us to represent the hopping parameter in the folowing form
\begin{widetext}
\begin{equation}
 \label{T(J)}
T(\rv,\rv+\av;t)\equiv T_0e^{iA(\rv,\rv+\av;t)} = 
\frac{\sqrt{4T_0^2|\rho(\rv,\rv+\av;t)|^2-J^2(\rv,\rv+\av;t)}+iJ(\rv,\rv+\av;t)}{2\rho(\rv,\rv+\av;t)},
\end{equation}
where the first term (the square root) in the enumerator is the real part of the complex link current, which is required to satisfy the constraint of Eq.~(\ref{T-modulus}). Equation (\ref{T(J)}) uniquely (mod $2\pi$) relates the link vector potential $A(\rv,\rv+\av;t)$ to the given physical link current $J(\rv,\rv+\av;t)$ and the many-body wave function $\psi(t)$. Inserting the hopping parameter of Eq.~(\ref{T(J)}) into Eq.~(\ref{SE2}) we obtain the following universal lattice NLSE
\begin{equation}
 \label{NLSE3}
i\partial_t\psi(\rv_1\dots\rv_N) = -\sum_{j,\av}
\frac{\sqrt{4T_0^2|\rho(\rv_j,\rv_j+\av)|^2-J^2(\rv_j,\rv_j+\av)}+iJ(\rv_j,\rv_j+\av)}{\rho(\rv_j,\rv_j+\av)}
\psi(\dots \rv_j+\av\dots)
 + \sum_{i>j}V_{\rv_i-\rv_j}\psi(\rv_1\dots\rv_N)
\end{equation}
\end{widetext}
that has to be solved with the initial condition of Eq.~(\ref{init-cond}) [we remind that the link density $\rho(\rv,\rv+\av;t)$ is defined by Eq.~(\ref{rho-def})]. The main existence and uniqueness results for the initial value problem of Eqs.~(\ref{NLSE3}) and (\ref{init-cond}), which control the existence of the lattice TDCDFT, are combined in the following theorem.

{\it Theorem 2. (Lattice TDCDFT)} --- Let the link currents $J(\rv,\rv+\av;t)$ be continuous functions of $t$, such that in the extended phase space ${\cal H}\times R$ there exists a subset $\Omega'\subset({\cal H}\times R)$ defined by the inequalities
\begin{equation}
 \label{Omega'}
2T_0|\rho(\rv,\rv+\av)| > |J(\rv,\rv+\av;t)|.
\end{equation}
If the initial point $(\psi_0,t_0)\in\Omega'$, then

\noindent
(i) There is a neighborhood of the initial point where the many-body wave function $\psi(t)$ is a unique retarded functional of $J(\rv,\rv+\av;t)$ and $\psi_0$, and the map $J\leftrightarrow A$ from the current to the vector potential is unique and invertible;

\noindent
(ii) The statement (i) can not be extended beyond some maximal existence time $t^*$, i.~e. the solution to Eqs.~(\ref{NLSE3}) and (\ref{init-cond}) is not global, if and only if at time $t^*$ the boundary of $\Omega'$ is reached. The latter means that at least for one link the local kinetic energy $|K(\rv,\rv+\av;t)|\to 0$ as $t\to t^*$.

{\it Proof} --- The proof of this theorem closely follows the proof of the Theorem~1 above. Indeed, the NLSE (\ref{NLSE3}) is formally a system of $N_{\cal H}$ first order ODE of the general form (\ref{abstract-NLSE}). The condition of Eq.~(\ref{Omega'}) guarantees that in Eq.~(\ref{NLSE3}) the expressions under all square roots are positive, and no one denominator vanishes. Therefore for all points $({\bm\psi},t)\in\Omega'$ the right hand side in Eq.~(\ref{NLSE3}) is a continuously differentiable function of $\psi$-variables with an explicit time dependence determined by the $t$-continuous link currents $J(\rv,\rv+\av;t)$. Hence if the initial point $(\psi_0,t_0)\in\Omega'$, the Picard-Lindel\"of theorem guarantees the local existence and uniqueness of a solution $\psi[J,\psi_0](t)$, which, in turn, implies the existence of the bijective map $J\leftrightarrow A$. 
Because the wave function, being normalized, is bounded, the only possibility for the solution $\psi[J,\psi_0](t)$ to be not global is that at some time $t^*$ it reaches the boundary of the subset $\Omega'$ where the right hand side of Eq.~(\ref{NLSE3}) is defined. In physical terms this means that at $t=t^*$ the local kinetic energy $K(\rv,\rv+\av;t^*)$ [the square root in Eq.~(\ref{NLSE3})] vanishes at least for one link. Otherwise the unique functional $\psi[J,\psi_0](t)$ and the bijective map $J\leftrightarrow A$ exist globally in time. End of the proof.

Apparently the conditions of the Theorem~2 are more restrictive than those in the Theorem~1. Because the left hand side of the inequality (\ref{Omega'}) is bounded, not all currents can be $A$-representable even locally, which should be contrasted to the unconditional local $T$-representability in the generalized TDCDFT. The physical reason for this is that the constraint of Eq.~(\ref{T-modulus}) fixes the hopping rate and thus does not allow for arbitrary large link currents. In virtue of the continuity equation (\ref{continuity2}) this also forbids arbitrary fast variations of the density, as has been noticed in Refs.~\onlinecite{Baer2008,LiUll2008,Verdozzi2008}. From Eq.~(\ref{Omega'}) one can easily estimate an upper bound on the admissible link currents. Indeed, by analogy with Ref.~\onlinecite{Verdozzi2008}, we can use the Cauchy-Schwarz inequality to find that
\begin{widetext}
\begin{equation}
\label{Schwarz}
|\rho(\rv,\rv+\av)|^2 
\le N^2\sum_{\rv_2\dots\rv_N}|\psi(\rv,\rv_2\dots\rv_N)|^2
\sum_{\rv_2\dots\rv_N}|\psi(\rv+\av,\rv_2\dots\rv_N)|^2 = n(\rv)n(\rv+\av),
\end{equation}
\end{widetext}
which leads to the folowing upper bound for $A$-representable link currents
\begin{equation}
 \label{upper-bound}
|J(\rv,\rv+\av)| < 2T_0\sqrt{n(\rv)n(\rv+\av)} \le \left\{
\begin{array}{ll}
 NT_0, & {\rm bosons}\\
 2T_0, & {\rm fermions}
 \end{array}
\right.
\end{equation}

In spite of the additional restrictions on the currents, the main outcome of the Theorem~2 is basically the same as for the generalized TDCDFT. If the condition (\ref{Omega'}) is satisfied at the initial time $t_0$, then any continuous link current is locally $A$-representable and the wave function is a unique functional of that current, which holds both for interacting and for noninteracting systems. Physically the condition (\ref{Omega'}) simply means that in the initial state the local link kinetic energy should be nonzero everywhere. In practice one almost always starts with the equilibrium/ground state with zero initial current. In this case the local uniqueness and existence ($A$-representability) are guaranteed. In fact, the standard setup for practical KS-based applications of TDCDFT is even less restrictive. What we actually need is a possibility to reproduce in the KS system the physical current $J$ flowing in the interacting system, where it is produced by some external potential $A^{\rm ext}$. Hence by construction $J$ is interacting $A$-representable. Theorem~2 states that at least locally (i.~e. during some finite interval of time) one can always find a selfconsistent potential $A^S=A^{\rm ext} + A^{\rm xc}[J]$ which will do the required job, provided we choose the initial state of the KS system with nonzero link kinetic energies. In general one can not exclude a situation when the physical for the interacting system current $J$ will eventually drive the noninteracting KS system to the boundary of the $A$-representability domain $\Omega'$. However this would mean that there is a link with exactly zero KS kinetic energy, which looks quite exotic. Therefore it is plausible to assume that in most physical situations the KS system should exist globally.

According to the Theorem~2 the possibility of having the same current in two different systems (say interacting and noninteracting) does not require any special relation between the initial states, except for nonvanishing local kinetic energies. However, the identical currents imply automatically only that the time-dependent parts of the densities in two systems are the same. If we want to guarantee the completely identical densities, we should impose an additional restriction on the admissible initial states. The initial states for two systems should yield the same initial density. If the dynamics starts from the ground state the best choice for the noninteracting initial state is, apparently, the KS ground state. 

I conclude this section by noting that, similarly to the case of the generalized TDCDFT considered in Sec.~III,  the NLSE of Eq.~(\ref{NLSE3}) is exactly integrable for a one-particle system ($N=1$). The corresponding solution is given by Eqs.~(\ref{modPsi-1part}) and (\ref{phase-1partQ}) with the local kinetic energy $K(\rv,\rv+\av;t)$ of the following form
\begin{eqnarray}
 \nonumber
&&K(\rv,\rv+\av;t) \\ \label{K-1particle}
 &=& \sqrt{4T_0^2|\psi(\rv;t)|^2|\psi(\rv+\av;t)|^2-J^2(\rv,\rv+\av;t)}
\end{eqnarray}
On can check by a direct substitution that the wave function $\psi(\rv;t)=|\psi(\rv,t)|e^{i\chi(\rv,t)}$, where $|\psi(\rv,t)|$ and $\chi(\rv,t)$ are determined by Eqs.~(\ref{modPsi-1part}), (\ref{phase-1partQ}), and (\ref{K-1particle}), indeed solves Eq.~(\ref{NLSE3}) for $N=1$. The explicit form of the functional $A[J,\psi_0](\rv,\rv+\av;t)$ is obtained by inserting this solution into Eq.~(\ref{T(J)}). 

The exact solution provides us with a useful illustration of the Theorem~2. Because of the presence of the square root in Eq.~(\ref{K-1particle}) the solution makes sense only if $4T_0^2n(\rv)n(\rv+\av)>J^2(\rv,\rv+\av)$ for all links. These are exactly the inequalities, which determine the subset $\Omega'$ in the theorem. If the prescribed current and the initial state are such that the above inequalities are never violated, the solution is global and given explicitly by  Eqs.~(\ref{modPsi-1part}), (\ref{phase-1partQ}), and (\ref{K-1particle}). The only alternative is that the solution, and thus the mapping, exists up to some finite time $t^*$, when at least for one link the local kinetic energy of Eq.~(\ref{K-1particle}) turns into zero, i.~e. the presupposed current becomes too large to be supported by adjusting only phases of the hopping parameters. The $N=1$ example also explicitly shows that if at $t=t_0$ the $A$-representability inequalities are fulfilled (the expression under the square root in Eq.~(\ref{K-1particle} is initially positive for all links), then for any $t$-continuous current the solution always exists, at least for some finite interval of time. The Theorem~2 states that for all interacting many-body systems the behavior, in a sense of the current-to-vector potential mapping and the $A$-representability, is qualitatively similar to the exactly solvable case of one particle.

\section{Conclusion}

In this work I have proved the existence of TDCDFT for lattice many-body systems. The proof is based on the observation that the density-to-potential mapping and the ${\cal V}$-representability problems can be reformulated in terms of a solution of a universal NLSE. For lattice systems the existence and the uniqueness of solutions to this NLSE can be proved by employing a well developed theory of nonliner ODE. 

To make the presentation more transparent I considered only lattices with nearest neighbor hopping, and ignored possible internal degrees of freedom, such as spin or the presence of more than one orbital on each site. These effects, as well as the spin-orbit interaction, can be included straightforwardly as they do not change the general mathematical structure of the many-body Schr\"odinger equation, which is essential for the present proof. Therefore the main statements of the Theorems~1 and 2 should be valid for all lattice many-body systems. 

From the practical point of view, the results of this work are also sufficient to justify the applications of TDCDFT, which rely on solving the Schr\"odinger equation on a real space grid. If we believe that the finite difference representation of the quantum many-body problem is an adequate approximation to the physical reality, then the TDCDFT is valid to the same level of accuracy. Of course, conceptually this solution of the problem is not satisfactory and an independent proof for the continuum version of the theory is need. The present work strongly suggests that the NLSE approach is the right way to go.

One of the key features of the present approach to TDCDFT is that it offers a direct access to the universal functional $A[J]$ via the solution of an explicit and well posed initial value problem. This may be helpful in studying the exact properties of the exchange-correlation potential in TDCDFT. In particular, it provides us with a straightforward procedure for a ``reverse engineering'' of the KS vector potential. For example, if for some simple/model interacting system we are able to solve exactly the interacting problem, then, using the corresponding current $J(t)$ as the input, we can almost immediately reconstruct the exact KS vector potential by solving the noninteracting version of Eq.~(\ref{NLSE3}), which is of the same level of computational complexity as solving the standard time-dependent KS equations.

\bigskip
\bigskip
This work was supported by
the Spanish MEC (FIS2007-65702-C02-01), ``Grupos Consolidados UPV/EHU del Gobierno
Vasco'' (IT-319-07), and the European Union through e-I3 ETSF project (Contract No. 211956).


\end{document}